\documentclass[usenatbib]{mn2e}

\usepackage[british]{babel}
\usepackage{amsfonts}
\usepackage{amsmath}
\usepackage{graphicx}
\usepackage{mathptmx}
\usepackage{dcolumn}
\usepackage{color,soul}
\usepackage{textcomp}

\begin{document}

\title[Imaging on PAPER: Centaurus A at 148 MHz]{Imaging on PAPER: Centaurus A at 148 MHz}
\author[Stefan et al.]{Irina I. Stefan,$^1$\thanks{E-mail: iis21@cam.ac.uk} Chris L. Carilli,$^2$ David A. Green,$^1$ Zaki Ali,$^{3}$ James E. Aguirre,$^4$ 
\newauthor Richard F. Bradley,$^{5,6,7}$ David DeBoer,$^{3}$ Matthew Dexter,$^{3}$ Nicole E. Gugliucci,$^7$ 
\newauthor D. E. Harris,$^8$ Daniel C. Jacobs,$^9$ Pat Klima,$^7$ David MacMahon,$^{3}$ Jason Manley,$^{10}$
\newauthor David F. Moore,$^4$ Aaron R. Parsons,$^{3,11}$ Jonathan C. Pober$^{11}$ and William P. Walbrugh$^{10}$ \\
$^1$Cavendish Laboratory, 19 J. J. Thomson Ave., Cambridge, CB3 0HE \\
$^2$National Radio Astronomy Observatory, Socorro, NM, USA \\
$^{3}$Radio Astronomy Laboratory, University of California, Berkeley, CA, USA \\
$^4$Department of Physics and Astronomy, University of Pennsylvania, Philadelphia, PA, USA \\
$^5$Department of Electrical and Computer Engineering, University of Virginia, Charlottesville, VA, USA \\
$^6$National Radio Astronomy Observatory, Charlottesville, VA, USA \\
$^7$Department of Astronomy, University of Virginia, Charlottesville, VA, USA \\
$^8$Smithsonian Astrophysical Observatory, Harvard-Smithsonian Center for Astrophysics, Cambridge, MA, USA\\
$^9$School of Earth and Space Exploration, Arizona State University, Tempe, AZ, USA \\
$^{10}$ Square Kilometer Array -- South Africa Project, Cape Town, South Africa\\
$^{11}$Astronomy Department, University of California, Berkeley, CA, USA}
\maketitle

\begin{abstract}
We present observations taken with the Precision Array for Probing the Epoch of Reionization (PAPER) of the Centaurus A field in the frequency range 114 to 188 MHz. The resulting image has a $25\arcmin$ resolution, a dynamic range of 3500 and an r.m.s. of 0.5 Jy beam$^{-1}$ (for a beam size of  $25\arcmin \times 23\arcmin$). A spectral index map of Cen A is produced across the full band. The spectral index distribution is qualitatively consistent with electron reacceleration in regions of excess turbulence in the radio lobes, as previously identified morphologically. Hence, there appears to be an association of `severe weather' in radio lobes with energy input into the relativistic electron population. We compare the PAPER large scale radio image with the X-ray image from the ROSAT All Sky Survey.  There is a tentative correlation between radio and X-ray features at the end of the southern lobe, some 200 kpc from the nucleus, as might be expected from inverse Compton scattering of the CMB by the relativistic electrons also responsible for the radio synchrotron emission. The magnetic fields derived from the (possible) IC and radio emission are of similar magnitude to fields derived under the minimum pressure assumptions, $\sim 1~\umu$G. However, the X-ray field is complex, with large scale gradients and features possibly unrelated to Cen A. If these X-ray features are unrelated to Cen A, then these fields are lower limits.

\end{abstract}

\begin{keywords}
galaxies: individual (Centaurus A) -- instrumentation: interferometers -- galaxies: active -- galaxies: structure -- X-rays: galaxies
\end{keywords}

\section{Introduction}

At a distance of 3.8 Mpc \citep{2010PASA...27..457H}, Centaurus A (Cen A) is by far the closest radio galaxy, showing AGN-driven radio jet structures on scales from parsecs to 200 kpc \citep{1998A&ARv...8..237I}.  Given its proximity, and very large physical scale, Cen A has become a key tool in the study of radio source physics. However, its large angular scale ($> 10\degr$) presents a severe challenge to radio imaging of extended features with reasonable spatial resolution.

The last few years have seen a renaissance in low frequency radio interferometry, driven principally by the goal of studying the redshifted \mbox{H\,{\sc i}} 21 cm emission from the neutral intergalactic medium during cosmic reionization over a redshift range of $6 < z < 14$, or frequency range 180 MHz to 100 MHz \citep{2006ARA&A..44..415F, 2006PhR...433..181F, 2010ARA&A..48..127M}. A number of radio interferometers are currently in operation to study reionization, including the Murchison Widefield Array (MWA; \citealp{2013PASA...30....7T}), the Low-Frequency Array (LOFAR; \citealp{2006astro.ph.10596R}), and the Donald C. Backer Precision Array for Probing the Epoch of Reionization (PAPER; \citealp{2010AJ....139.1468P}). In parallel to the reionization studies, the combination of good sensitivity, reasonable spatial resolution, and very wide field of view, of these interferometers presents a perfect opportunity to image large-scale, non-thermal, radio continuum structures, such as those of Cen A.

Here we present the imaging results on Cen A from the PAPER array in South Africa. PAPER, with a field of view of $\sim 60\degr$ and spatial resolutions between $\sim 15\arcmin~\rm{and}~25\arcmin$ across the band, is ideally suited to image the extended emission from Cen A. In Section \ref{Centaurus_A} we give a brief overview of relevant studies of Cen A. In Sections \ref{Observations} and \ref{Data reduction} we review the observational techniques and describe the reduction of the PAPER data. We make a detailed comparison with the higher resolution morphological analysis of Cen A at 1.4 GHz by Feain et al. (2011) in Section \ref{Results}. Section \ref{spectral index} presents the spectral index map of Cen A across the PAPER bandwidth. We go on in Section \ref{Analysis} to compare the radio and X-ray structures, and show that the spectral behaviour of Cen A is consistent with particle acceleration in turbulent structures in the radio lobes. Finally, Section \ref{Conclusion} summarizes our conclusions.

\section{Centaurus A}\label{Centaurus_A}

Centaurus A is considered the archetype of Fanaroff--Riley type I radio galaxies (i.e. edge-darkened, lower luminosity radio galaxies -- usually with luminosities at $178$~MHz below $2\times 10^{25} ~\rm{W}~\rm{Hz}^{-1}\rm{sr}^{-1}$, \citealp{1974MNRAS.167P..31F}). \cite{1998A&ARv...8..237I} offers a comprehensive review of Cen A up to its publication year. 

Cen A is part of the Centaurus A/M83 Group of galaxies. The host galaxy, NGC 5128, is a massive elliptical galaxy ($M=4.1 \times 10^{11}~M_{\odot}$; \citealp{1976ApJ...208..673V}), with a prominent, kpc-scale dust-lane. This dust lane likely indicates a recent merger with a dusty spiral galaxy, as is also suggested by the presence of optical shells and filaments around the host galaxy \citep{1983ApJ...272L...5M, 1994ApJ...423L.101S}. Stellar kinematics in the nuclear region of Cen A indicate a mass of the black hole of $M_{BH} = (5.5 \pm 3.0) \times 10^7~M_{\odot}$ \citep{2009MNRAS.394..660C}.

At radio frequencies, early observations of Cen A revealed two giant outer lobes, extending up to 260 kpc north and south of the core, and possibly further \citep{1965AuJPh..18..589C}. Subsequent imaging has delineated radio structures from sub-parsec scale jets \citep{1998AJ....115..960T}, through two well defined inner lobes and jets on scales of $\sim$5 kpc \citep{1983ApJ...273..128B}, as well as a `middle lobe' in the northern part of the source on a scale of 30 kpc \citep{1999MNRAS.307..750M}.

The most comprehensive radio imaging study of the outer lobes of Cen A to date was by \cite{2011ApJ...740...17F}. They combined observations by the Australia Telescope Compact Array and the Parkes 64-m radio telescope to image Cen A at $1.4$ GHz with a spatial resolution of $\sim 50\arcsec$ over a scale of $10\degr$. In this image they observe a number of new features in the outer lobes. The hook-like structure at the northern-most end of the northern lobe terminates in a series of radially-centric shells with decreasing brightness. These shells have similar curvature to a diffuse filament protruding from the eastern edge of the lobe, prompting \cite{2011ApJ...740...17F} to suggest that both are modelled by the local group `weather' \citep{2009MNRAS.393....2S}. Near the northeastern edge of the hook they observe a partial ring and suggest that was created by the passage of a neighbouring galaxy. In the southern lobe, \cite{2011ApJ...740...17F} identify a `vertex' and a `vortex' close to  $13^{\rmn{h}}~20^{\rmn{m}}$ and $-45\degr~15\arcmin$, which they suggest could be a region of plasma reawakened by strong shocks arising from Cen A's core or by the passage of a dwarf galaxy. They also remark on the detection of a turbulent Faraday rotation measure signal \citep{2009ApJ...707..114F} from the southern lobe and the presence of surface wave-like structures \citep{1990ApJ...357..373B} close to the western edge of the lobe and suggest that, together with the `vertex' and a `vortex', they could be the signature of Kelvin--Helmholtz instabilities. They account for the `vertex' and a `vortex' being part of the wave-like structure by conjecturing a backflow dominated southern lobe with turbulent mixing.

A detailed comparison of the X-ray and radio emission on scales of order 10~kpc shows clear evidence for the interaction between the expanding radio source and the thermal cluster gas \citep{2002ApJ...577..114K}. On much larger scales ($\sim200$~kpc), early X-ray imaging by the Ariel V satellite showed possible evidence for extended emission from the outer radio lobes \citep{1978MNRAS.182..661C}. \cite{1978MNRAS.182..661C} suggest the emission is inverse-Compton upscattering of the CMB by the relativistic electrons in the radio lobes, and calculate a radio lobe magnetic field of $0.7~\umu$G, and that this field is comparable to the equipartition field. \cite{1994A&A...288..738A}, presents the ROSAT sky survey X-ray emission on a scale of $10\degr$ around Cen A. This image shows a marginal excess of X-ray emission from the end of the northern radio lobe. \cite{2012arXiv1210.4237S} present the latest X-ray analysis of Cen A with the Suzaku satellite. Their spectra of lobe regions show possible diffuse emission consistent with a thermal gas in the radio lobes with a temperature $0.5$~keV and density $\sim 10^{-4}$~cm$^{-3}$, although the very large scales involved and the relative proximity of Cen A to the Galactic plane (situated 10$\degr$ to 20$\degr$ off the Galactic plane) lead to some confusion in the X-ray image. If real, this thermal gas has a pressure comparable to that derived from minimum energy conditions. We return to the X-ray-radio comparison below.

Cen A has also been studied in the $> 100~\rm{GeV}$ energy range and its radio lobes are considered a likely origin for ultra-high energy cosmic rays \citep{2012ApJ...753...40F}.

\section{Observations}\label{Observations}

PAPER is an interferometric transit array aimed at detecting the fluctuations in the $21$~cm emission from the Epoch of Reionization. The project currently consists of a 32-antenna test-array situated at NRAO, in Green Bank, USA, and a 64-antenna array in South Africa's Radio Quiet Zone (latitude of approximately $30\degr$ S). Further reference to PAPER here implies only the South African array. The antennas are non-steerable, operate within the frequency range of 110--190~MHz and have dual linear polarization capability. The primary beam is $60\degr$ FWHM with a spatially and spectrally smooth response down to the horizon \citep{2012AJ....143...53P}.

We present here data taken in single polarization (E--W) on four consecutive nights (2011 July 3\textsuperscript{rd}/4\textsuperscript{th} -- 6\textsuperscript{th}/7\textsuperscript{th}) from approximately 4pm to 8am each `night'. This observing campaign offers various investigation opportunities, such as studying the Galactic Centre or searching for radio transients and supernova remnants. Here we present the imaging results on Cen A. All the coordinates used in this paper are in J2000.

\begin{figure}
	\centerline{\includegraphics[width=\columnwidth]{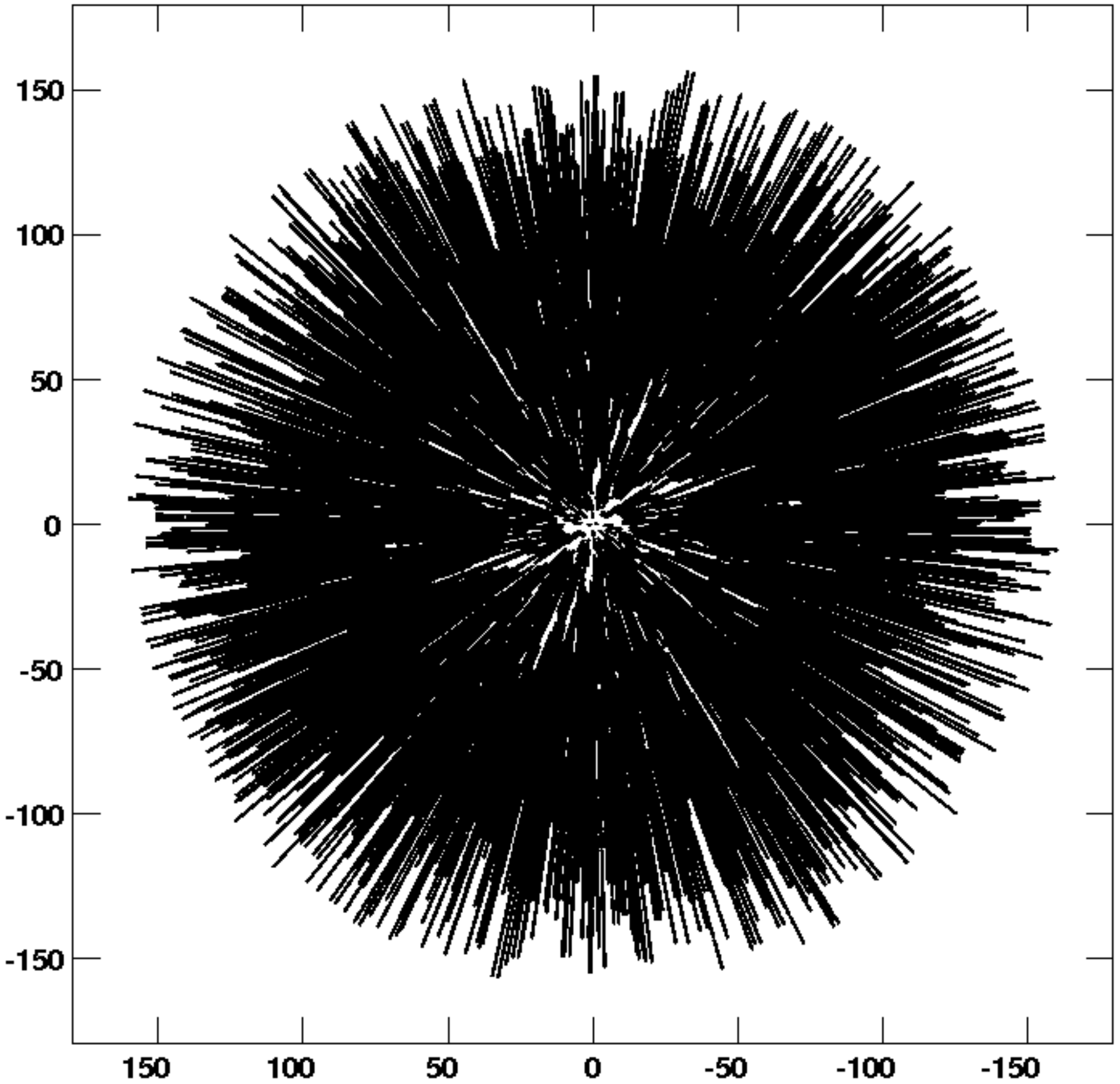}}
	\caption{PAPER $uv$-coverage for the full band and a 10-min dataset. The axes are labelled in units of wavelength.}
	\label{coverage}
\end{figure}

At the time, the 64 antennas of the array were deployed in a minimum redundancy configuration \citep{2012ApJ...753...81P} within a 300 m-wide circle, with an empty central circular region of radius 10 m. This generates resolutions between $\sim 15\arcmin~\rm{and}~25\arcmin$ across the band. Visibilities were integrated and recorded every $10.7$~s. The full bandwidth is divided into 2048 narrow channels, each of width $48.83$~kHz. Since the antennas are not steerable, the observations were taken as a drift-scan of the sky. They were approximately continuous and divided into datasets of about $10$-min long. Fig. \ref{coverage} shows the $uv$-coverage across the full PAPER band for one 10-min dataset.

Only datasets in which Cen A was within approximately 1.2 hours from meridian transit are considered. Between meridian transit and the outermost position the primary beam gain drops by about $32\%$. Table \ref{data_table} shows which time intervals were used from each night for Cen A (about 2 hours) and the number of $10$-min datasets used.

\begin{table}
\caption{Time intervals for Cen A observations.}
\label{data_table}
\begin{tabular}{c c c c} 
\hline
Date & \multicolumn{2}{c}{Time interval}  & Number of \\
& (UT) & (s) & datasets \\
\hline
2011 Jul 3\textsuperscript{rd} & 16:12:35 -- 18:32:53 & 8418 & 15 \\
2011 Jul 4\textsuperscript{th} & 15:54:57 -- 18:25:16 & 9019 & 16 \\
2011 Jul 5\textsuperscript{th} & 16:09:09 -- 18:09:25 & 7216 & 13 \\
2011 Jul 6\textsuperscript{th} & 15:46:19 -- 18:16:38 & 9019 & 16 \\
\hline
\end{tabular}
\end{table}

\section{Data reduction}\label{Data reduction}

Initial pre-calibration was performed using the Astronomical Interferometry in PYthon (\textsc{AIPY}) package\footnote{http://casper.berkeley.edu/astrobaki.index.php/AIPY}. This corrects hardware and header errors and subtracts cross-talk. An initial manual Radio Frequency Interference (RFI) excision step of the most prominently damaged channels was performed (mainly a few communication satellites bands, e.g. Orbcomm near $137$~MHz). We find that due to PAPER's location, RFI only affects a few narrow channel ranges. Frequencies below 114~MHz and above~188 MHz were flagged due to bandpass roll-off. Within each $10$-min dataset, we phase tracked the mean zenith, such that the result is similar to that of a tracking interferometer and can be imaged as such.

The main calibration was carried out using the Astronomical Image Processing System (\textsc{AIPS}) package\footnote{http://www.aips.nrao.edu/index.shtml}. A second round of manual RFI excision was performed by inspecting the scalar averages for the total power spectra for each dataset and flagging outlying peaks. Approximately $67\%$ of the initial data remained after both RFI excision steps. A final clipping at 3 times the average visibility level ensures that any remaining outliers are removed.

Each $10$-min dataset was divided into $15$ broad frequency channels (4.98~MHz wide subsets) of 102 narrow channels to mitigate bandwidth smearing effects. The $10$-min time length of a dataset is sufficiently short for the effect of the source moving through PAPER's primary beam to be negligible. All subsequent steps only used baselines longer than 5 wavelengths, as the shortest baselines appear to be affected by residual low level interference and very large scale structures which are not well sampled due to PAPER's non-uniform $uv$-coverage (see Fig. \ref{coverage}) at very short baselines.

Given the field of view of $\sim60\degr$ of PAPER, the observed field cannot be approximated as a flat patch of sky and the three-dimensionality of the visibility function needs to be taken into account. The sky brightness distribution is recovered as the surface of a unit sphere embedded in the three-dimensional Fourier transform of the visibility function. \cite{1999ASPC..180..383P} discusses the effect of non-coplanar baselines on wide observing fields, as well as some proposed solutions. However, PAPER is a two-dimensional array and each $10$-min dataset is sufficiently short for the observation to be regarded only as projections of the sphere onto a succession of tangent planes given by the geometry of the array at the time of each observation. For each `snapshot' there is a change in the apparent positions of the sources away from the tangent point. From \cite{1999ASPC..180..383P}, a source (Cen A) at declination $\sim-40\degr$, $r \sim 0.3~\rm{rad}~(1.2$ hours) away from the phase centre will have a positional offset of approximately $30\arcmin$. This is comparable to the restoring beam used for imaging ($25\arcmin \times 23\arcmin$), so the only effect to the final image is a slight E--W broadening of Cen A. The rest of the sources in the field will experience a variable amount of broadening as well, correlated to their distance from the phase centre. For the flux calibrators, we took this into account when finding their flux densities.

The datasets were self-calibrated to obtain complex gain solutions for each antenna. Those with gain variations higher than $30\%$ were flagged. Only four antennas had such high and unstable gains to warrant removal from at least one of the nights. The following parameters were found to give good solutions after two phase self-calibration loops and a single phase and amplitude one: a solution interval of 2 minutes, a minimum number of 5 antennas allowed for a solution and a minimum allowed signal to noise ratio of 4. Clean components were used until the first negative.

Each subset was \uppercase{CLEAN}ed \citep{1980A&A....89..377C} down to a set average flux level. This was taken to be the average root mean square noise (r.m.s.) of all the 10-min data subsets for a given night and channel range. Imaging was performed with a restoring beam of size of $25\arcmin \times 23\arcmin$, the lowest resolution across the band. A \uppercase{ROBUST} factor of $-1$ was used to define the $uv$-weighting scheme (a factor of $-4$ is nearly pure uniform weighting, while a factor of $+4$ is nearly pure natural weighting).

The images were normalized by the primary beam model as a function of frequency as in the survey by \cite{2011ApJ...734L..34J}, and resampled onto the same sky-grid, using the datatset closest to meridian tranzit as reference. For each broad channel, the images corresponding to different time intervals were averaged together. We measured the flux densities of bright sources within $18\fdg5$ from zenith for each channel range. This radius was chosen such that Cen A is included in the area thus defined. We manually identified these sources with ones from the Culgoora catalogue \citep{1995AuJPh..48..143S}, eliminated the ones with multiple inconsistent measurements for the same frequency and obtained 13 sources to act as flux density calibrators. We fitted spectra to the historical data for each of these sources with quadratic functions. The flux density calibrators, their positions and their derived spectral parameters are listed in Table \ref{flux_cal}. A least squares fit was performed for each channel range in order to derive scaling factors for the maps based on the calibrators' measured and expected spectra.

\begin{table*}
\caption{The 13 sources used as flux calibrators with their Culgoora name and positions and the derived spectral fit parameters. The functional form $\log_{10}S = A + B \times \log_{10}\nu + C \times \log_{10}^{2}\nu$, with S measured in Jy and $\nu$ in MHz.}
\label{flux_cal}
\begin{tabular}{c c c D{.}{.}{4.2} D{.}{.}{4.2} D{.}{.}{4.2} D{.}{.}{4.2}}
\hline
Culgoora & RA & Dec & \multicolumn{1}{c}{$A$} & \multicolumn{1}{c}{$B$} & \multicolumn{1}{c}{$C$} & \multicolumn{1}{c}{Flux at}\\
name & & & & & & \multicolumn{1}{c}{148 MHz (Jy)}\\
\hline
Cul 1355$-$416 & $13^{\rmn{h}}~55^{\rmn{m}}~57.4^{\rm{s}}$ & $-41\degr~38\arcmin~14\arcsec$ & $8.0$ & -$0.8$ & $0$ & 25.9\\ 
Cul 1308$-$220 & $13^{\rmn{h}}~08^{\rmn{m}}~57.6^{\rm{s}}$ & $-22\degr~00\arcmin~33\arcsec$ & -$16.5$ & $5.0$ & $0.3$ & 67.3\\
Cul 1407$-$425 & $14^{\rmn{h}}~07^{\rmn{m}}~23.5^{\rm{s}}$ & $-42\degr~32\arcmin~56\arcsec$ & $15.9$ & -$2.6$ & $0.1$ & 10.9\\
Cul 1413$-$364 & $14^{\rmn{h}}~13^{\rmn{m}}~32.3^{\rm{s}}$ & $-36\degr~26\arcmin~55\arcsec$ & $1.6$ & $0.5$ & -$0.1$ & 13.3\\ 
Cul 1424$-$418 & $14^{\rmn{h}}~24^{\rmn{m}}~45.9^{\rm{s}}$ & $-41\degr~52\arcmin~58\arcsec$ & $16.9$ & -$3.3$ & $0.2$ & 8.2\\
Cul 1421$-$382 & $14^{\rmn{h}}~21^{\rmn{m}}~11.6^{\rm{s}}$ & $-38\degr~13\arcmin~29\arcsec$ & $9.0$ & -$1.0$ & 0 & 16.9\\
Cul 1419$-$272 & $14^{\rmn{h}}~19^{\rmn{m}}~54.5^{\rm{s}}$ & $-27\degr~14\arcmin~13\arcsec$ & $22.2$ & -$4.0$ & $0.2$ & 21.9\\
Cul 1422$-$297 & $14^{\rmn{h}}~22^{\rmn{m}}~32.9^{\rm{s}}$ & $-29\degr~47\arcmin~06\arcsec$ & $13.5$ & -$2.1$ & $0.1$ & 17.0\\
Cul 1206$-$337 & $12^{\rmn{h}}~06^{\rmn{m}}~06.0^{\rm{s}}$ & $-33\degr~47\arcmin~26\arcsec$ & -$3.2$ & $1.9$ & -$0.2$ & 12.8\\
Cul 1218$-$373 & $12^{\rmn{h}}~18^{\rmn{m}}~01.4^{\rm{s}}$ & $-37\degr~23\arcmin~51\arcsec$ & -$8.2$ & $2.8$ & -$0.2$ & 8.0\\
Cul 1221$-$423 & $12^{\rmn{h}}~21^{\rmn{m}}~04.0^{\rm{s}}$ & $-42\degr~19\arcmin~30\arcsec$ & -$17.5$ & $4.6$ & -$0.3$ & 11.0\\
Cul 1232$-$416 & $12^{\rmn{h}}~32^{\rmn{m}}~58.7^{\rm{s}}$ & $-41\degr~36\arcmin~39\arcsec$ & -$9.6$ & $3.2$ & -$0.2$ & 9.7\\
Cul 1252$-$289 & $12^{\rmn{h}}~52^{\rmn{m}}~00.7^{\rm{s}}$ & $-28\degr~57\arcmin~13\arcsec$ & $5.6$ & $0$ & -$0.1$ & 14.7\\
\hline
\end{tabular}
\end{table*}

\section{Results}\label{Results}

\begin{figure*}
	\centerline{\includegraphics[width=\textwidth,angle=0,trim= 0cm 0cm 0cm 0cm, clip=true]{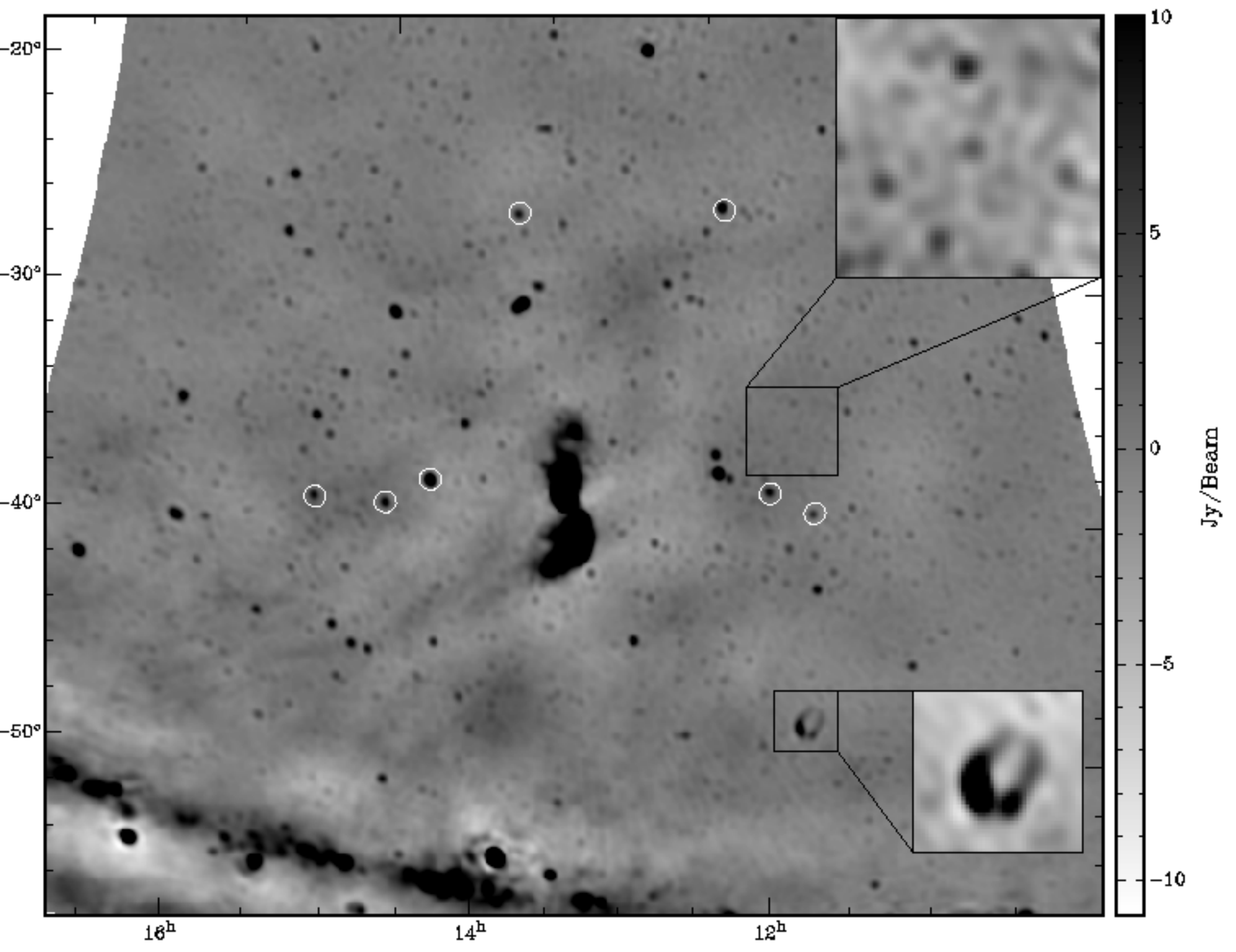}}
	\caption{PAPER final image of Centaurus A obtained with a $25\arcmin \times 23\arcmin$ FWHM Gaussian restoring beam. Sources listed in Table \ref{fluxes} are marked with white circles. The two magnified patches are on different greyscales to each other and to the general image. The lower magnified patch shows the SNR PKS 1209$-$52.}
	\label{Cena_field}
\end{figure*}

Fifteen calibrated maps were produced, one for each channel range. Fig. \ref{Cena_field} shows a weighted average of the middle thirteen. It has a dynamic range of $\sim 3500$ as found from the ratio of the peak brightness of the core of Cen A to the r.m.s. of $0.5$~Jy $\rm{beam}^{-1}$ of a region near the edge of the image. Cen A dominates the centre of the image and the Galactic Plane crosses it further South. The Supernova Remnant (SNR) PKS 1209$-$52 is visible to the SW of Cen A, at $-52\degr$~Dec, as a shell with two brightened limbs \citep{1976AuJPh..29..435D}. Table \ref{fluxes} lists the flux densities measured from Fig. \ref{Cena_field} for a number of sources in the Cen A field. Flux densities extrapolated from Culgoora measurements are also listed for comparison. There is reasonable agreement for all but one of the sources, Cul 1334$-$296. We looked at the historical observational data for the extended source Cul 1334$-$296 (M83) and found that there is a high degree of variation of measured flux densities. Nevertheless, we extrapolated a flux density at 150 MHz from all but the Culgoora measurements and obtained 11.1 Jy, higher than the Culgoora expected flux density. There is the possibilty of a turnover point occuring at frequencies higher than the Culgoora measurements, but given that there are frequencies in the historical observational data for Cul 1334$-$296 with two mutually inconsistent measurements, we do not believe this alternative to be highly likely and mantain confidence in our measurement of its flux density.

\begin{table}
\caption{Observed flux densities for some sources in the Cen A field as found with the \uppercase{jmfit} task in \textsc{AIPS} from an average map of the middle thirteen channel ranges. The expected flux density is found from the Culgoora 160 MHz measurements and the corresponding spectral indices.}
\label{fluxes}
\begin{tabular}{c D{.}{.}{4.2} D{.}{.}{4.2} D{.}{.}{4.2}} 
\hline
Culgoora & \multicolumn{1}{c}{Expected} & \multicolumn{1}{c}{Observed flux} & \multicolumn{1}{c}{R.m.s.} \\
source & \multicolumn{1}{c}{flux density} & \multicolumn{1}{c}{density} & \multicolumn{1}{c}{(Jy $\rm{beam}^{-1}$)} \\
name & \multicolumn{1}{c}{(Jy)} & \multicolumn{1}{c}{(Jy)} & \\
\hline
Cul 1221$-$423 & 8.1 & 5.1 & 0.7 \\
Cul 1232$-$416 & 11.0 & 9.2 & 0.7 \\
Cul 1252$-$289 & 14.9 & 16.2 & 0.7 \\
Cul 1334$-$296 & 2.8 & 8.4 & 0.7 \\
Cul 1355$-$416 & 29.5 & 33.5 & 0.7 \\
Cul 1407$-$425 & 9.6 & 10.2 & 0.7 \\
Cul 1424$-$418 & 9.9 & 9.9 & 0.7 \\
\hline
\end{tabular}
\end{table}

\begin{figure*}
	\centerline{\includegraphics[angle=-90, width=\textwidth, trim= 1cm 3cm 1cm 2cm, clip=true]{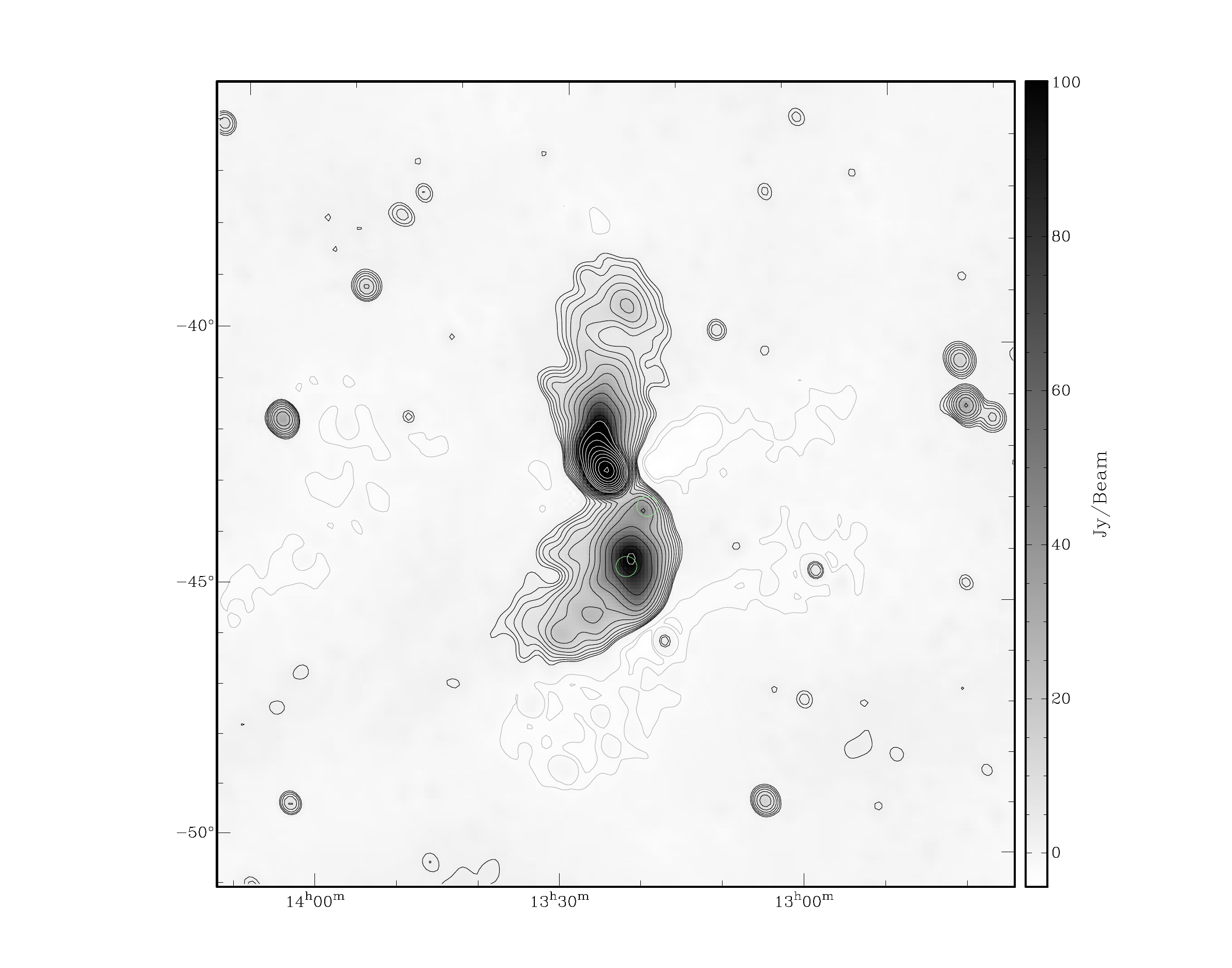}}
	\caption{PAPER image of Centaurus A obtained with a $25\arcmin \times 23\arcmin$ FWHM Gaussian restoring beam. Contour levels are a geometric progression in $\sqrt{2}$, staring from $2$~Jy beam$^{-1}$. Negative contours are in light grey. The green circles mark the positions of MRC 1318--434B and PKS 1320--446.}
	\label{Centaurus A}
\end{figure*}

A lower limit of 5 wavelengths was imposed for calibration and imaging, which means structures larger than $11\fdg5$ are not imaged. The negative dips in the maps evince the lack of short spacing data. Cen A is surrounded by a shallow region of negative flux. This is even more visible in and around the Galactic plane, since the Galaxy has structures on effectively all spatial scales. Again, these negative `bowls' represent missing short-spacings in the interferometric images, and result in uncertainty in the total flux measurements of large sources, and diffuse surface brightnesses. Hence, the spectral index discussion in Section \ref{spectral index} will be mostly qualitative.

However, there is some evidence that Galactic emission extends all the way to Cen A (see the end of this section) and large-scale foreground emission will certainly have a significant Galactic contribution. From \cite{1970AuJPA..16....1L}, the brightness temperature in the direction of Cen A at 150~MHz is 360~K. Assuming an angular area of 14 deg$^2$, the galactic foreground contribution over Cen A will be approximately 1200~Jy. So, in a way, the $11\fdg5$ angular cutoff acts as a filter for part of this emission.

Fig. \ref{Centaurus A} is a magnified view of Cen A. The core and inner lobes are saturated to show the structure in the outer lobes more clearly. The peak brightness is found near $13^{\rmn{h}}~25^{\rmn{m}}~31^{\rm{s}}$ and $-43\degr~01\arcmin~26\arcsec$ to be $1745$ Jy $\rm{beam}^{-1}$. The flux density over the entire source (taking the lowest contour level as the extent-limit for the lobes) is $4100$ Jy at the average frequency of 148 MHz of the map. \cite{1981A&A...100..209H} found the flux density of Cen A at 408 MHz to be $2762 \pm 370$ Jy. Assuming an average spectral index over the entire source of $-0.7$ between 19.7 MHz and 960 MHz \citep{1960PASP...72...29B}, then the expected flux density at 148 MHz is 5600 Jy. \cite{1958AuJPh..11..400S} measured Cen A's flux density at 85.7 MHz to be $8700 \pm 1300$ Jy. Extrapolating again to 148 MHz gives 5900 Jy. Both of these values are significantly higher than the one measured here, stressing the importance of the missing short spacings, but neither \cite{1981A&A...100..209H} nor \cite{1958AuJPh..11..400S} correct their observations for Galactic contributions, which, again, would amount to $\sim 1200$ Jy. The average depth off-source of the negative bowl is $-2.7$~Jy beam$^{-1}$; assuming an angular area of 14 deg$^2$ and with a restoring beam of $25\arcmin \times 23\arcmin$, the negative bowl accounts for approximately $250$~Jy. Including these two effects brings the PAPER measurement within error of the extrapolated values for the \cite{1981A&A...100..209H} and \cite{1958AuJPh..11..400S} measurements.

The northernmost part of the southern lobe displays emission from an extended background source, the radio galaxy MRC 1318--434B \citep{1990PKS...C......0W, 1986ApJ...308...36S}, associated with NGC 5090. This can be seen as a small bright patch on the northern edge of the southern lobe. There is a second bright background radio source in the centre of the southern lobe, PKS 1320--446 \citep{1990PKS...C......0W, 2006AJ....131..114B}, but at our resolution we estimate that this source is at most 30\% of the flux density measured over a single synthesized beam, and does not affect the overall analysis below.

The southern lobe shows a sharp edge along the southwestern boundary of the radio source, and more diffuse emission extending to the east in two `prongs' of emission by about 2$\degr$. These two prongs are also seen at 1.4~GHz \citep{2011ApJ...740...17F}. Also seen in this image is the `gap' between the southern lobe and the nucleus, where the brightness drops below 7 Jy beam$^{-1}$ (as opposed to an average of 15 Jy beam$^{-1}$ on the northern edge of the southern lobe). This gap is also seen at 1.4~GHz \citep{2011ApJ...740...17F}.

The northern middle lobe extends outwards from the inner lobe and curves northward into the outer lobe, such that the outer lobe has a position angle of approximately $0\degr$. At about $13^{\rmn{h}}~28^{\rmn{m}}$ and $-40\degr~33\arcmin$, the outer lobe curves east to form a hook-like structure. The northern lobe shows a general decline in brightness away from the core, with further brightening in the hook-like region at the northern end of the radio lobe, some 260 kpc from the core.

On the largest scales, faint emission may extend a few degrees beyond the two-prongs to the southeast, although missing short spacings start to become an important issue in this regard. This very extended emission to the southeast has been noted previously, and may relate to high-latitude Galactic emission, and not Cen A \citep{1960PASP...72...29B, 1981A&A...100..209H}.

\section{C\lowercase{en} A spectral index map}\label{spectral index}

\begin{figure}
	\centerline{\includegraphics[angle=0, width=\columnwidth, trim= 0cm 0cm 0cm 0cm, clip=true]{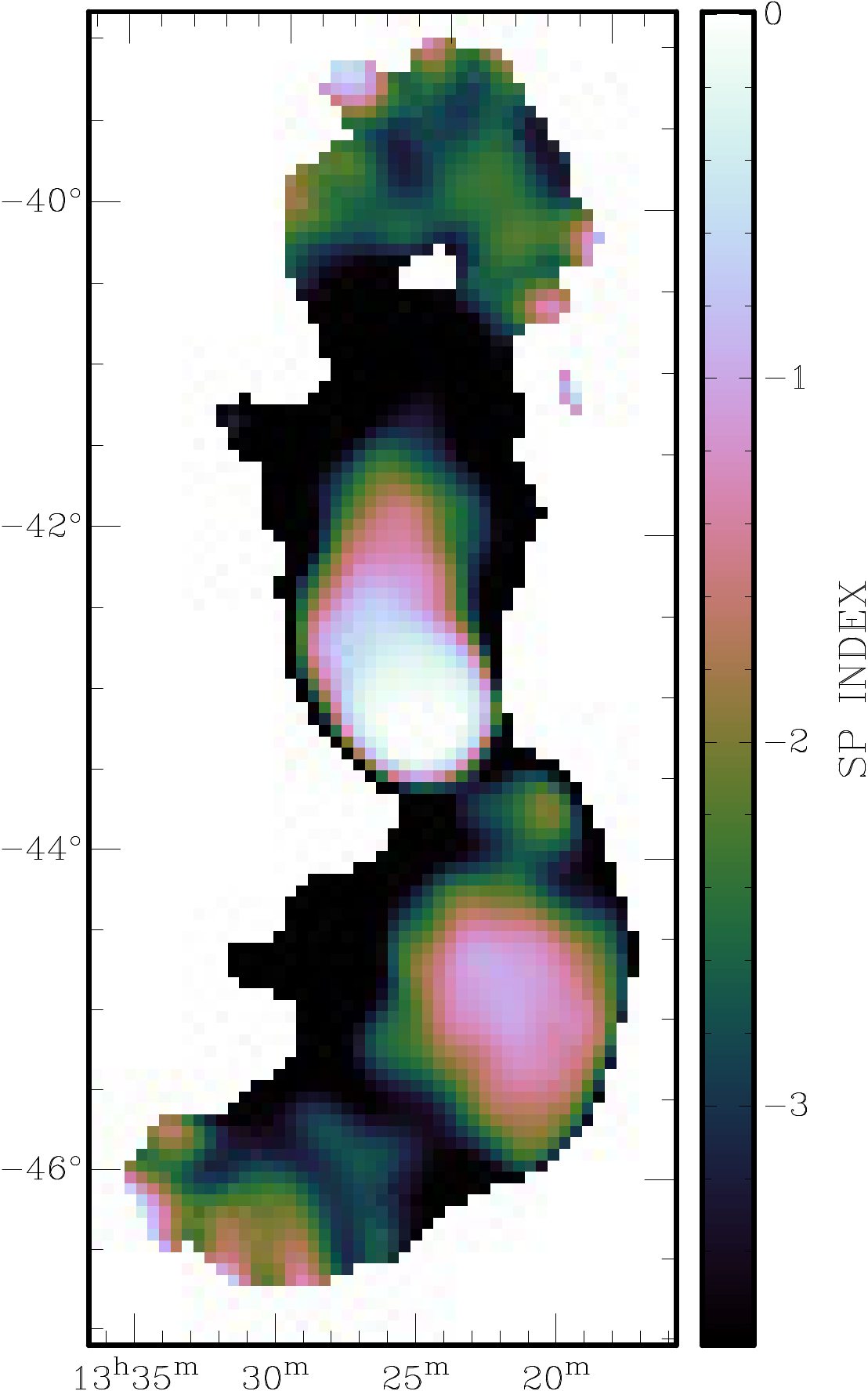}}
	\caption{Spectral index map of Cen A. The resolution is $25\arcmin$. The `cubehelix' \citep{2011BASI...39..289G} colour scheme is used.}
	\label{spindex_map}
\end{figure}

PAPER's large bandwidth allows for spectral index maps of the observations to be computed across the band. The spectral index\footnote{The convention $S \propto \nu^{\alpha}$ is used.} image is shown in Fig. \ref{spindex_map}. The spectral index was calculated at each point between a weighted average image of 4 lower frequency channels (123.58 MHz -- 138.38 MHz) and a weighted average of 4 higher frequency channels (158.11 MHz -- 172.90 MHz). All 8 images were created using a restoring beam of $25\arcmin \times 23\arcmin$. The r.m.s. of each image is used as a weight during averaging. Both average images were blanked at a $5\sigma$ level. As mentioned before, incomplete reconstruction of the total flux will bias this calculation, but the small scale trends are still valid.

The northern lobe shows the spectral steepening over the first 3$\degr$ from the radio core, characteristic of FR I (edge-darkened) radio galaxies. Interestingly, the spectral index then becomes flatter in the northern hook-like structure, furthest from the core. 

The southern lobe also shows interesting spectral behaviour. The centre of the lobe and the southwestern hard-edge show flatter spectral indices, with steepening toward the more diffuse prong to the east. And again, the very end of the southern-most prong shows flattening of the spectrum towards its tip.

The background radio galaxy on the northern edge of the southern lobe, MRC 1318--434B, can be seen as an unresolved flatter point in the spectral index map.

\section{Analysis}\label{Analysis}

\begin{figure}
	\centerline{\includegraphics[width=\columnwidth,angle=0,trim= 0cm 0cm 0cm 0cm, clip=true]{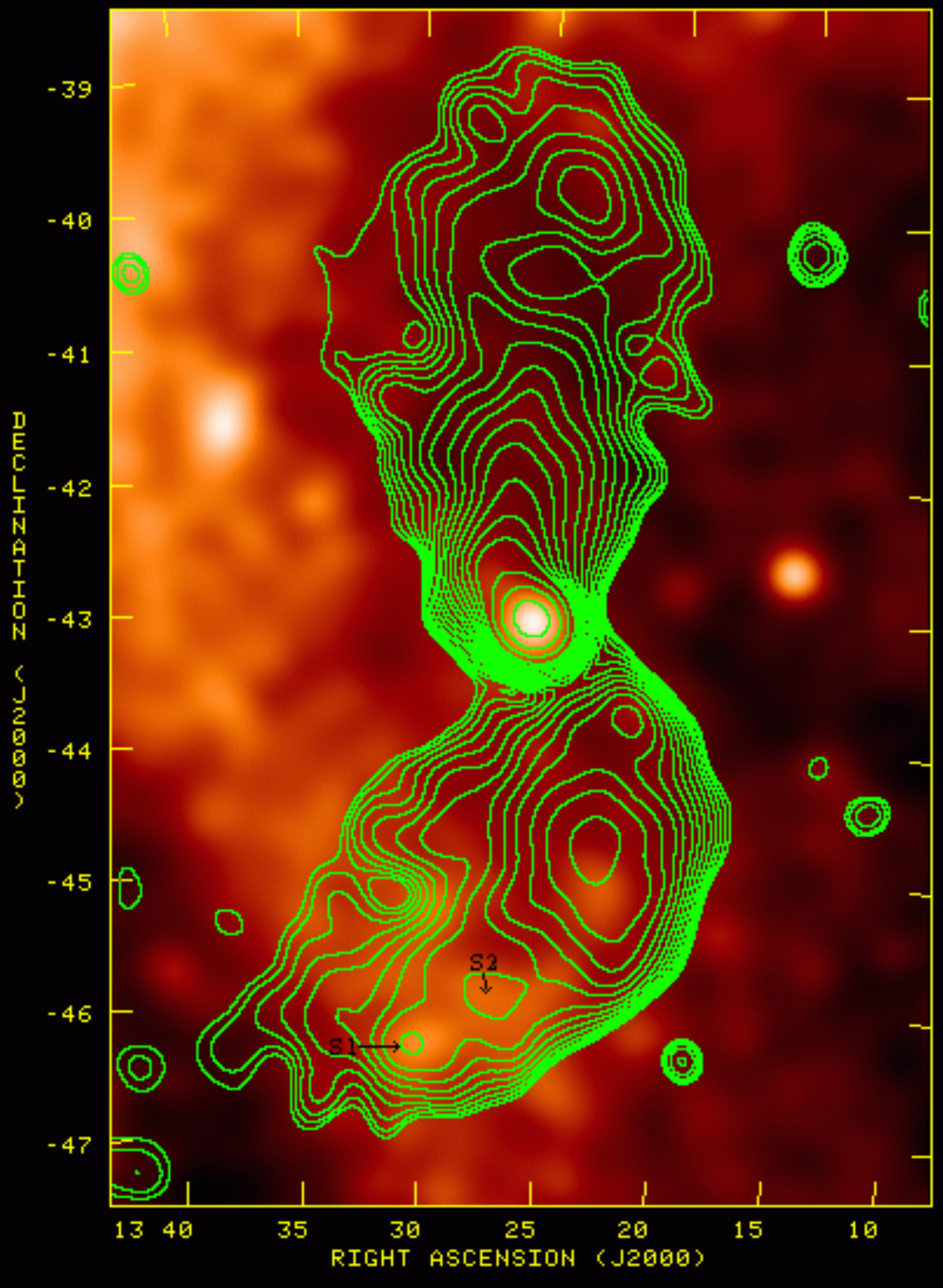}}
	\caption{X-ray image of Cen A, taken from the ROSAT all-sky survey and smoothed with a $25\arcmin \times 23\arcmin$ beam. Overlaid in green are the radio contours from 1Jy up to 100 Jy.}
	\label{Xray}
\end{figure}

\subsection{Spectral Index}

Spectral flattening in radio lobes is typically interpreted as a sign of energy input into the relativistic electron population \citep{1991ApJ...383..554C}. While necessarily qualitative due to the short spacing issue, a comparison of the spectral index distribution and the morphological structures seen at high resolution, and their interpretation by \cite{2011ApJ...740...17F}, is illustrative.

Starting with the southern lobe, we find that the central region, and the sharp southwestern edge, correspond to flatter spectral indices than the regions to the east (and close to the core, although this region is confused by a background radio source). This region is called-out explicitly in the \cite{2011ApJ...740...17F} analysis as being rich in high contrast, turbulent structure, notably their `vortex', `vertex', and `waves'.  

Likewise, the end of the northern lobe was highlighted by \cite{2011ApJ...740...17F} as exhibiting multiple shells and a ring in the 1.4 GHz image. This region also shows substantial spectral flattening relative to the regions closer to the core.

\subsection{ROSAT All Sky Survey Comparison}\label{ROSAT}

Comparison of X-ray and radio emission in radio galaxies can be a powerful tool for probing the physics of the radio source and its interaction with the surrounding medium. Possible signatures include: (i) X-rays arising from inverse Compton scattering of the ambient photons by the relativistic electrons also responsible for the radio synchrotron emission \citep{1979MNRAS.188...25H, 1980ARA&A..18..165M}, and (ii) an anti-correlation between non-thermal radio emitting plasma and thermal X-ray emitting cluster gas due to the hydrodynamic interaction between the radio lobes and the ambient medium \citep{2007ARA&A..45..117M, 2012ARA&A..50..455F}.

The difficulty with Cen A is the very large angular scale for the outer lobes.  The only available X-ray image comparable in coverage to the PAPER radio image is the ROSAT All Sky Survey\footnote{www.xray.mpe.mpg.de/cgi-bin/rosat/rosat-survey} (RASS; \citealp{1999A&A...349..389V}). In Section \ref{Centaurus_A} we reviewed the previous X-ray and Radio analyses of Cen A, including a possible spectroscopic detection of thermal X-ray emission in the vicinity of the radio lobes \citep{2012arXiv1210.4237S}.

Fig. \ref{Xray} shows an image of the $0.1$ -- $2.4$ keV emission from the RASS with the PAPER Cen A contours overlaid. The X-ray image is dominated by a large scale filament extending roughly N--S, first pointed out by \cite{1994A&A...288..738A}. \cite{2012arXiv1210.4237S} speculate that this filament may be Galactic in origin, although there is no direct evidence as to the location of the filament (Galactic or extragalactic).

A number of radio and X-ray correlations can be seen. The most prominent is the core and inner jet of Cen A, on scales $< 1\degr$, or $< 50$ kpc. This X-ray emission on the scale of the Cen A host galaxy has been well studied by \cite{2002ApJ...577..114K}. A second, fainter coincidence is seen at the position of the galaxy NGC 5090 at the northern edge of the southern lobe.

There are also a series of local peaks in the X-ray image in the southern half of the southern lobe, starting near the peak of the radio emission and extending southwest. These knots roughly align with the curving ridge of emission seen in the radio. We note that these X-ray knots are only marginally brighter than other knots in the large-scale X-ray filament in this vicinity, but outside the radio lobe, and hence this structure could simply be coincidence.

If these X-ray features relate to the radio structures, then they may be due to inverse Compton upscattering of the ambient CMB photons (the dominant photon field in the outer lobe regions) by the synchrotron emitting relativistic electrons \citep{1979MNRAS.188...25H, 1980ARA&A..18..165M}. Under this assumption, one can derive the magnetic field strengths by comparing the radio and X-ray luminosities from a given region.  We determine the X-ray luminosities after convolving the counts to $15\arcmin$ resolution, and subtracting a background based on an annulus with the same area as the photometric circle. The resulting X-ray luminosities and magnetic fields are given in Table \ref{B_field}. We find fields between 0.1 $\umu$G and 1.3 $\umu$G, depending on the radio spectral index. If the X-ray knots are unrelated to the radio emission, then these fields are lower limits since a weaker field would lead to stronger X-ray emission (i.e. require more relativistic electrons to explain the radio emission).

For comparison, we calculate the magnetic field strength in these regions assuming minimum pressure conditions \citep{1980ARA&A..18..165M}, with a relativistic electron to proton ratio of $k=1$, unity filling factor, upper and lower frequency cutoffs for the radio spectrum of 10 MHz and 10 GHz and a luminosity distance of 3.75 Mpc. We assume that the emission is coming from a spherical region in the lobe of $15\arcmin$ radius. The resulting fields are given in Table \ref{B_field}. The values range from 0.5 $\umu$G to 1 $\umu$G, depending on the spectral index assumed. The fields based on a minimum pressure assumption are of similar magnitude to those derived based on inverse-Compton emission assuming the X-rays are from the radio lobes.

\cite{2009MNRAS.393.1041H} find an upper limit for the density of the X-ray emitting gas in Cen A's lobes of $10^{-4}~\rm{cm}^{-3}$, which agrees with the range derived by \cite{2012arXiv1210.4237S}. Assuming $n_{g} = 10^{-4}~\rm{cm}^{-3}$ and a gas temperature of $kT \sim 0.5$~keV (as found by \citealp{2012arXiv1210.4237S}), then the lobes' thermal pressure is approximately $8 \times 10^{-14}~\rm{erg}~\rm{cm}^{-3}$. \cite{2012arXiv1210.4237S} found that this thermal pressure counterbalances the pressure of a non-thermal plasma ($8 \times 10^{-14}~\rm{erg}~\rm{cm}^{-3}$) with an average magnetic field of strength $B \sim 1~\umu$G in rough equipartition with ultra-relativistic electrons. 

More sensitive, very wide field X-ray imaging is required to test the reality of these X-ray features, and their possible association with the radio source in Cen A.

\begin{table}
\caption{Magnetic field strengths found from the RASS X-ray image. $L_{\rmn X-ray}$ is the measure X-ray luminosity, $B_{minP}$ the magnetic field strength found through the minimum pressure condition and $B_{IC}$ the magnetic field strength found through assuming inverse Compton scattering. S1 and S2 are marked in Fig. \ref{Xray} and $\alpha$ is the spectral index. We use two values for the spectral index to account for the uncertainty in the absolute $\alpha$ measured from Fig. \ref{spindex_map}.}
\label{B_field}
\begin{tabular}{c c c c c} 
\hline
&\multicolumn{2}{c}{S1} &  \multicolumn{2}{c}{S2} \\
&$\alpha = -1$&$\alpha = -1.8$&$\alpha = -1$&$\alpha = -1.8$\\
\hline
log($L_{\rm{X-ray}}(\rm{erg}~\rm{s}^{-1}$)) & 39.0 & 39.2 & 39.9 & 40.1 \\
$B_{\rm{minP}}$ ($\umu$G) & 0.56 & 0.80 & 0.67 & 0.96 \\
$B_{\rm{IC}}$ ($\umu$G) & 0.25 & 1.25 & 0.12 & 0.74 \\
\hline
\end{tabular}
\end{table}

\section{Conclusions}\label{Conclusion}

We have presented wide field imaging of large scale structures taken with the Precision Array for Probing the Epoch of Reionization in the frequency range 114 to 188 MHz of the closest radio galaxy, Centaurus A, and a comparison to the large scale X-rays. 

We observe a correspondence between regions rich in complex morphological radio structure at 1.4 GHz (described by \citealp{2011ApJ...740...17F}) and regions of flatter spectral index in the PAPER map centered on 148 MHz. The northern loop of the northern lobe, where \cite{2011ApJ...740...17F} show a series of shells and a ring, sees a flattening of the spectral index relative to the regions closer to the core.  Similar flattening is seen in the central and western part of the southern lobe, where \cite{2009ApJ...707..114F}  observe a `vertex' and a `vortex'. This likely indicates active particle acceleration on large scales in the radio lobes, through supersonic turbulences. We note that the gap between the core of the galaxy and the northern edge of the southern lobe, which has also been observed at 1.4 GHz \citep{2011ApJ...740...17F}, is still seen at 148 MHz. 

Considering the RASS comparison, the end of the southern lobe shows a possible correlation of X-ray emission with a series of radio knots seen in the PAPER radio image. Again, given the presence of the large scale X-ray filament, these features could simply be a coincidence in
the complex X-ray field. However, if these features are related to the radio emission, then an inverse Compton model leads to magnetic fields of the order of $1~\umu$G, comparable to those derived assuming a minimum pressure configuration for the relativistic fields and particles.

\section*{Acknowledgments}\label{Acknowledgments}
We thank the SKA project office in South Africa for their efforts in ensuring the smooth running of PAPER.
The PAPER project is supported through the NSF-AST program (awards 0804508, 1129258, and 1125558), and by significant efforts by staff at NRAO's Green Bank and Charlottesville sites. We acknowledge grant support from the Mt. Cuba Astronomical Foundation.
We have made use of the ROSAT Data Archive of the Max-Planck-Institut fur extraterrestrische Physik (MPE) at Garching, Germany.
We thank the referee for hepful comments.

\setlength{\labelwidth}{0pt}

\end{document}